\documentclass[psf,epsf,twocolumn,showpacs,preprintnumbers,floatfix]{revtex4}
\usepackage{graphics}
\usepackage{graphicx}
\usepackage{dcolumn} 
\usepackage{bm}
\usepackage{epsfig}
\newcommand{\sto}{SrTiO$_3$}
\newcommand{\lto}{LaTiO$_3$}

\newcommand{\tith}{Ti$^{3+}$}
\newcommand{\tif}{Ti$^{4+}$}

\begin{document}
\title{Correlation-Driven Charge Order at a \\ Mott Insulator - Band
Insulator Digital Interface}
\author{Rossitza Pentcheva$^{1}$} 
\email{pentcheva@lrz.uni-muenchen.de}
\author{Warren E. Pickett$^{2}$} 
\affiliation{$^{1}$Department of Earth and Environmental Sciences, 
University of Munich, Theresienstr. 41, 80333 Munich, Germany}
\affiliation{$^{2}$Department of Physics, 
  University of California, Davis, California 95616}
\date{\today}
\begin{abstract}
To study digital Mott insulator LaTiO$_3$  and band insulator
SrTiO$_3$ interfaces, we apply correlated band
theory (LDA+U) to (n,m) multilayers, $1\leq n,m \leq 9$.
If the on-site repulsion on Ti is large enough to 
model the magnetic insulating behavior of cubic bulk \lto, 
the charge imbalance at the interface
is found in all cases to be accommodated by disproportionation
(Ti$^{4+}$ + Ti$^{3+}$), charge ordering, and Ti$^{3+}$ 
$d_{xy}$-orbital ordering, with antiferromagnetic
exchange coupling between the spins in the interface layer.  
Lattice relaxation affects the conduction behavior
by shifting (slightly but importantly) the lower Hubbard band, 
but the disproportionation
and orbital ordering are robust against relaxation.
\end{abstract}
\maketitle


Atomically abrupt (``digital'') interfaces (IFs) between oxides 
with strongly differing electronic properties  
(superconducting-ferromagnetic; ferroelectric-ferromagnetic) have 
attracted interest\cite{ivan,ijiri} 
due to the new behavior that may arise, and for
likely device applications.
Hwang and collaborators~\cite{ohtomo2002,hwang2006} have reported coherent
superlattices containing a controllable number of Mott insulator [\lto~(LTO)] 
and band insulator [\sto~(STO)] layers 
using pulsed laser deposition, with analysis suggesting atomically sharp 
interfaces comparable to those produced by molecular beam epitaxy~\cite{ivan}. 
The most provocative result was that the IFs of these insulators showed 
metallic conductivity and high mobility.
Electron energy loss spectra (EELS) for Ti 
suggested  a superposition of \tith~and \tif~ions in the interface region. 
Incorporating doping with magnetic ions, these same materials are being
explored for spin-dependent transport applications~\cite{herranz}.
Effects of structural imperfections are being 
studied~\cite{hwang2004,shibuya,muller2004,nakagawa}, but the ideal IFs
need to be understood first.

Single- and three-band Hubbard models with screened intersite Coulomb
interaction have been applied to this IF. Both the Hartree-Fock approximation or dynamical mean field with a semiclassical
treatment of correlation~\cite{SO_AJM} result in a ferromagnetic (FM)
metallic IF over a substantial parameter range.
{\it Ab initio} studies reported so far 
have focused on charge profiles~\cite{satpathy,hamann} while neglecting
correlation effects beyond the local density approximation (LDA) that
we address below.  Very recent {\it ab initio} calculations of the
effects of lattice relaxation~\cite{hamann,spaldin06} 
at the IF have provided additional input
into the Hubbard-modeling of these IFs, allowing the investigation
of the interplay of correlation effects and relaxation in these models.

\begin{figure}[b]
\begin{center}
\rotatebox{0.}{
\includegraphics[scale=0.48]{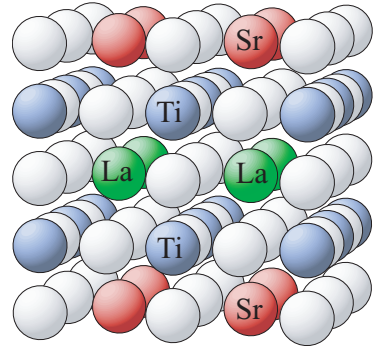}}
\end{center}
\caption{\label{structure}  Segment of the (1,1) LaTiO$_3$-SrTiO$_3$
 multilayer, illustrating the cubic perovskite structure
  (unlabeled white spheres denote oxygen).  An LaO layer lies
  in the center, bordered by two TiO$_2$ layers, with a SrO layer at top
  and bottom.  The lateral size of this figure corresponds to the
  $p(2 \times 2)$ cell discussed in the text.
    }
\end{figure}
The material-specific insight into correlated behavior that can
be obtained from first-principles-based approaches is still lacking.
In LTO/STO superlattices, the transition metal ions on the perovskite
B-sublattice are identical (Ti) and only the charge-controlling A-sublattice
cations (Sr, La) change across the interface ({\sl cf.} Fig.~\ref{structure}).
This leaves at each IF a TiO$_2$ layer whose local 
environment is midway between that
in LTO and STO.
In this paper we study
mechanisms of charge compensation at the LTO/STO-IF
based on density-functional theory calculations
[within the generalized gradient approximation
(GGA)~\cite{pbe96}] employing the
all electron FP-LAPW-method within the WIEN2k-implementation~\cite{wien}
including a Hubbard-type on-site Coulomb repulsion (LDA+U)~\cite{anisimov93}.
We focus on (1) the local
charge imbalance at the IF and its dependence on neighboring layers, (2)
the breaking of three-fold degeneracy of the Ti $t_{2g}$ orbitals 
which will be at most
singly occupied, (3) magnetic ordering and its effect on gap formation, and
(4) how rapidly the insulators heal (both in charge and in magnetic order)
to their bulk condition away from the IF.

To explore the formation of possible charge disproportionated, magnetically 
ordered,
and orbitally selective phases at the IF and to probe the relaxation
length towards bulk behavior we have
investigated a variety of ($n,m$) heterostructures with $n$ LTO and $m$
STO layers ($1 \leq n,m \leq 9$), and lateral cells of $c(2\times 2)$
or $p(2\times 2)$~\cite{details}. 
The $\hat z$ direction is taken perpendicular to the IF. 
Lattice parameters of the systems have been set to the experimental 
lattice constant of STO, 3.92~\AA, therefore modeling
coherent IFs on an STO substrate.
Bulk STO is a 
semiconductor with a GGA-band gap~\cite{pbe96} of 2.0 eV 
(experimental value 3.2~eV), separating
filled O $2p$ bands from empty Ti $3d$ bands.
  
Currently LTO ($a$=3.97 \AA)
and other $3d^1$ perovskites are intensively 
studied because their structure is crucial in determining their electronic 
and magnetic behavior~\cite{pavarini}.  Bulk LTO is an AFM insulator of 
G-type (rocksalt spin arrangement) with a 
gadolinium orthoferrite (20 atom) structure; however, lattice imaging
indicates that only a few layers of LTO assume the cubic structure that
we use in our superlattices.
Using the LDA+U method, an AFM insulator is obtained for
$U \geq 6$ eV, with a magnetic moment 
$M_{Ti} \approx 0.75 \mu_B$ due to
occupation of one of the $t_{2g}$ orbitals (orbital ordering arising from
spontaneous symmetry breaking).  
FM alignment of spins is 50 meV/Ti less favorable. 
\begin{figure}[b]
\hspace{-3mm}
\includegraphics[scale=0.78]{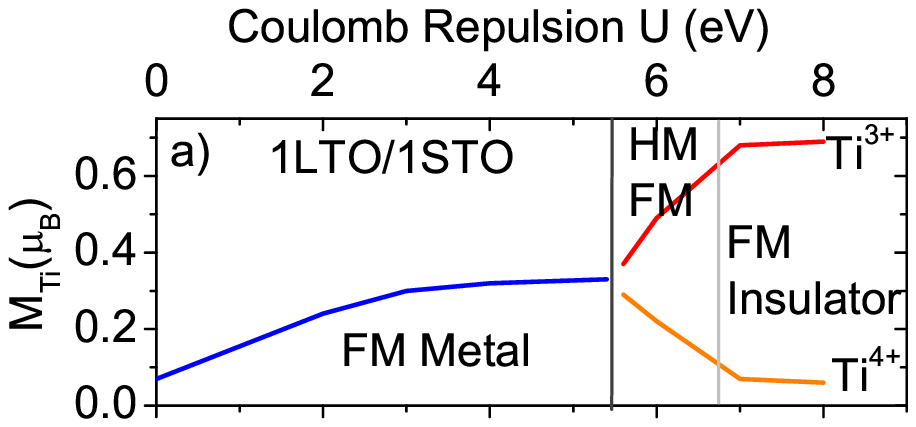}
\rotatebox{90.}{\includegraphics[scale=0.75]{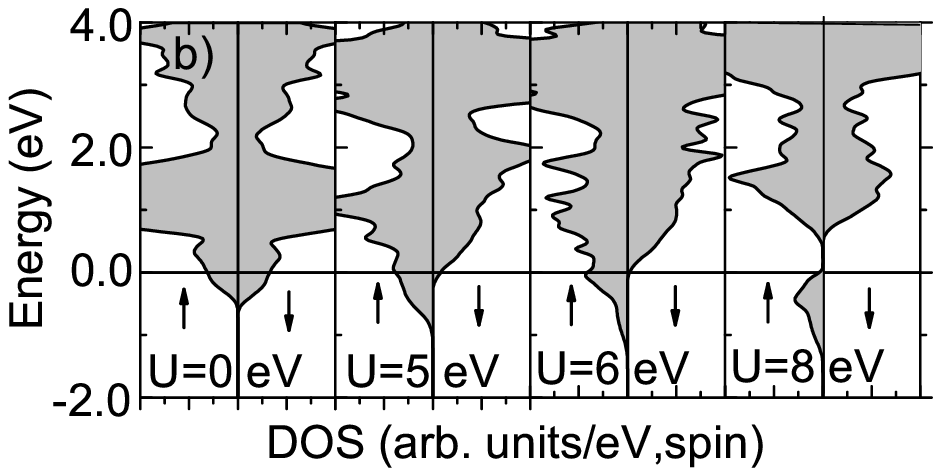}}
\caption{\label{MofU} a) Phase diagram of the Ti moments for
 the (1,1) superlattice in a
 transverse $c(2\times2)$ cell, versus the on-site Coulomb repulsion strength 
$U$ on the Ti $3d$ orbitals.  b) Density of states 
(spin direction indicated by arrows)
of the (1,1) superlattice for different $U$ values.
Disproportionation occurs in a weak first-order manner around
$U\approx$5.5 eV.
HM FM indicates a region of half 
metallic ferromagnetism before the Mott gap appears around $U \approx$ 6.5 eV.
}
          \end{figure}
We discuss first the (1,1) multilayer (1LTO/1STO layer) 
pictured in Fig. \ref{structure},
and then consider systems with thicker LTO and/or STO slabs to analyze 
the relaxation towards bulk behavior.

{\it The (1,1) superlattice}
is modeled in a transverse $c(2\times 2)$ cell 
(not considered in earlier work~\cite{spaldin06}) with two 
inequivalent Ti ions, which allows disproportionation within a 
single Ti layer and is consistent with 
the AFM G-type order in bulk LTO.  The on-site repulsion strength
$U$ on Ti was varied from 0 to 8~eV to assess both weak and strong 
interaction limits.
The Ti moment versus U and the evolution of the density of states as a 
function of U are shown in Fig. \ref{MofU}a) and b), respectively.  
Within GGA ($U=0$) nonmagnetic metallic character is obtained, consistent
with earlier reports.~\cite{satpathy,hamann} For 
$U\leq5$ eV
the system is a ferromagnetic metal with equivalent Ti ions, {\it i.e.}
it is qualitatively like earlier results on multiband Hubbard 
models.~\cite{SO_AJM}  At $U\approx$ 5.5 eV
disproportionation occurs on the Ti
ions, apparently weakly first-order as has been found to occur 
in the Na$_x$CoO$_2$ 
system~\cite{kwlee}.
Around $U \approx$ 6 eV there is 
a half metallic ferromagnet region, but beyond $U \approx$ 6.5 eV a gap 
opens separating  the lower Hubbard band and  resulting in a correlated 
insulator phase.
In the following we model the Mott insulating gap (0.5 eV) with $U$=8 eV.

\begin{figure}[t]
 \vspace{-5mm}
    \begin{center}
\includegraphics[scale=0.43]{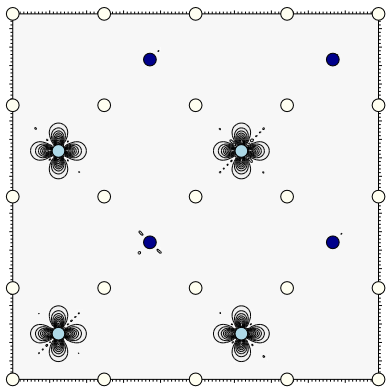}
    \end{center}
 \vspace{-10mm}
\caption{\label{CDNCOOO} 45$^{\circ}$ checkerboard
charge density distribution of the occupied $3d$ states in the 
charge-ordered TiO$_2$ layer in the FM (1,1) multilayer. Orbital-ordering
due to $d_{xy}$ orbital occupation is apparent.
The positions of O-, \tith~and \tif-ions are marked
by white, dark blue (black) and light blue (grey) circles, respectively.
}
\end{figure}

The arrangement of disproportionated ions, which is
charge-ordered (CO) rocksalt, retains
inversion symmetry and, more importantly, the more highly 
charged $d^0$ ions avoid being nearest neighbors.  
The spatial distribution of the occupied $d$-orbitals 
in the IF TiO$_2$ layer displayed in Fig.~\ref{CDNCOOO} reveals that 
besides the CO for $U>7$~eV this state is orbitally ordered (OO)
with a filled $d_{xy}$ orbital at the \tith~sites, the non-degenerate member 
of the cubic $t_{2g}$ triplet after the intrinsic symmetry-lowering effect 
of the IF. The Fermi level lies in a 
small Mott gap separating the occupied narrow $d_{xy}$ band 
(`lower Hubbard band') from the rest of 
the unoccupied $d$-orbitals.   For ferromagnetic alignment of the spins
($M_{\rm Ti^{3+}}=0.72\mu_{\rm B}$) the gap ensures an integer moment
($2.0\mu_{\rm B}$). 

The system at this level of treatment is a realization of a 
quarter-filled extended Hubbard model (EHM)
system.  The Hubbard model itself is metallic at quarter-filling; when 
intersite repulsion is included~\cite{onari,zhang} it becomes CO and insulating.
The intersite repulsion is included correctly in first principles methods
and that combined with the on-site repulsion ($U$) gives charge ordering.

The calculation was extended to a larger
$p(2\times2)$-cell 
to allow antiferromagnetic 
alignment of the \tith~spins.  We obtain the same 
CO/OO state
with an occupied $d_{xy}$-orbital on every second IF Ti ion, giving
a checkerboard ordering of \tith~and ~\tif, regardless of whether
the spins are aligned or antialigned.  
AFM coupling is preferred by 80~meV per 
$p(2\times2)$-cell for the (1,1) superlattice (a spin-spin exchange
coupling of $|J|$=10 meV). For heterostructures  
containing a thicker LTO slab, however, AFM coupled spins on the 50\% diluted  
p(2$\times$2) mesh in the IF layer will not match the AFM G-type order on 
the LTO side of the slab, where spins in the IF-1 layer couple antiparallel 
with a c(2$\times$2)-periodicity. Due to this frustration, AFM alignment 
within the IF layer may become less favorable.  
\begin{table}
\caption{\label{tab:MU_nm} Layer resolved magnetic moments (in 
$\mu_B$) of the Ti ions in ($n,m$) superlattices. 
Due to the $c(2\times 2)$-lateral unit cell there are  two inequivalent 
Ti-ions in each layer. $(n,m)$ denotes a multilayer containing $n$ LTO and $m$ 
STO layers.  The IF moments are nearly bulk-like and become so at the layer
next to the IF layer. (1,5)$^*$ denotes a configuration where the interlayer 
distances were relaxed according to Ref.\cite{spaldin06}.}
\begin{ruledtabular}
\begin{tabular}{cccccc}
  System & \multicolumn{2}{c}{LTO} & IF & \multicolumn{2}{c}{STO}  \\
   $(n,m)$ &IF-2 & IF-1 & IF  & IF+1 & IF+2 \\
\hline
(1,1)& - & - & 0.72/0.05 & - & - \\
(1,5) &      -     &     -      & 0.71/0.05 & 0.0/0.0 & 0.0/0.0 \\
(1,5)$^*$ &      -     &     -      & 0.50/0.08 & 0.0/0.0 & 0.01/0.01 \\
(5,1)& 0.73/-0.73 & -0.73/0.73 & 0.70/0.05 &    -    &    -    \\
(5,5)& 0.73/-0.73 & -0.73/0.73 & 0.70/0.06 & 0.0/0.0 & 0.0/0.0 \\
\end{tabular}
\end{ruledtabular}
\end{table}
\begin{figure}[b]
    \begin{minipage}{4. cm}
 \scalebox{0.85}{\includegraphics{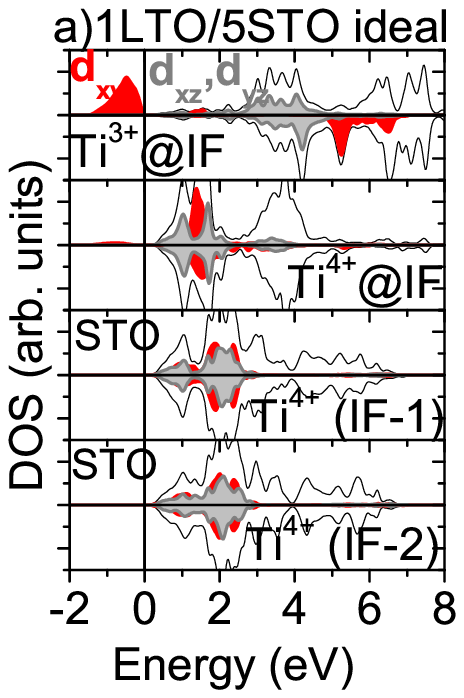}}
    \end{minipage}
    \begin{minipage}{3.5 cm}
 \hspace*{-.5cm}
  \scalebox{0.84}{\includegraphics{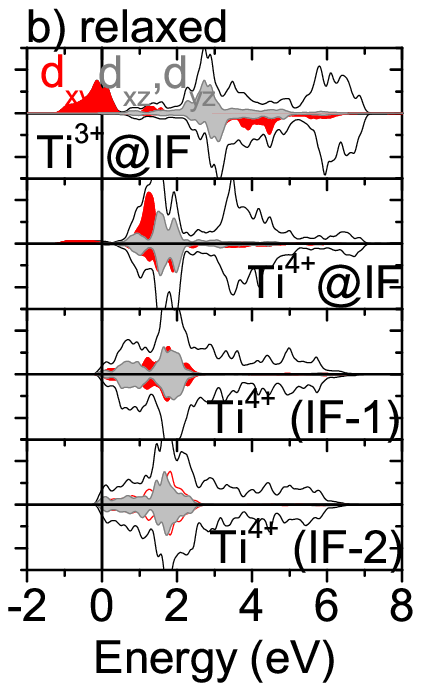}}
    \end{minipage}

\caption{\label{DOS55} Layer-resolved density of states of a 
a) structurally ideal and b) relaxed (1,5) multilayer. 
The two topmost panels show \tith~ and \tif~ at the IF, the succeeding panels
show the behavior of the Ti-ion in deeper layers of the STO part of the slab.
While for the ideal geometry rapid relaxation of the electronic structure to
bulk form versus distance from the interface (IF) takes place, in the relaxed 
structure the electronic relaxations involve deeper lying layers. 
}
\end{figure}

{\it The (n,m) superlattice.}
To examine charge- and spin-order relaxation towards bulk behavior,
and observe charge accommodation at more isolated IFs, we have studied several
thicker slabs containing $(n,m)$ layers of LTO and STO, respectively, with 
$1\leq n,m \leq 9$.  Following the experiment of Ohtomo  
{\it et al.}~\cite{ohtomo2002} we present results specifically for
the (1,5) and (5,1) as 
well as the (5,5) superlattices, all of
which we find to be disproportionated, CO and OO, and insulating in the strong 
interaction regime.
As is clear both from the layer-resolved magnetic moments 
presented in Table~\ref{tab:MU_nm} and the layer resolved projected 
DOS for the  (1,5) superlattice in Fig.~\ref{DOS55}a), 
the IF TiO$_2$ layer, and only this layer, is CO/OO with \tith~and 
\tif~distributed in a checkerboard manner. 
At every second Ti-site the $t_{2g}$ states 
split according to the IF-imposed symmetry lowering,
and the $d_{xy}$ orbital becomes occupied. 
The $t_{2g}$ states on the \tif~ions remain essentially 
degenerate, and there is only
a tiny induced moment $M_{\rm Ti^{4+}}= 0.06\mu_{\rm B}$.
Ti ions in neighboring or  deeper
layers in the STO part of the slab have the configuration 
$3d^0$ and are nonmagnetic, 
while those on the LTO side of the slab 
have the configuration $3d^1$ and are AFM G-type ordered. 
Thus the charge mismatch is localized at the interface 
layer, with bulk LTO and  STO character quickly re-emerging on 
neighboring layers. Consequently, these results indicate a relaxation length 
much less that the 1-2~nm value
estimated from the EELS data~\cite{ohtomo2002}. 
The same CO/OO results have been obtained on a variety of $(n,m)$ 
LTO-STO superlattices; 
these repeatedly emerging insulating ordered IF phases are very robust.

However, the systems discussed so far are structurally 
perfect with ideal positions of the atoms in the perovskite lattice. 
In the following we discuss the influence of lattice relaxations 
on the electronic 
properties of the system. Recently, two DFT studies using GGA~\cite{hamann} 
and the LDA+U approach~\cite{spaldin06} investigated structural 
relaxations in \lto/\sto~superlattices, finding that Ti-ions at the IF 
are displaced by 0.15\AA\ with respect to the oxygen ions leading to a 
longer Ti-Ti distance through 
the LaO layer than through the SrO-layer. This ``ferroelectric''-like 
distortion decays quickly in deeper lying layers. Using the relaxations 
reported in Ref.\cite{spaldin06}, we repeated the calculations for the 
(1,5)-heterostructure. The resulting layer-resolved projected DOS at the 
Ti-ions is displayed in Fig.~\ref{DOS55}b). The most prominent feature is that 
for the relaxed structure the $d_{xy}$-band (the lower Hubbard band) has 
been shifted up by 0.4 eV,  
leaving it incompletely ($\sim$70\%)  occupied. 
The charge is distributed in the minority spin channel at the \tith-sites 
(hybridization with O$2p$ bands) reducing the magnetic moment from 
$0.71\mu_{\rm B}$ in the ideal structure to $0.50\mu_{\rm B}$. Additionally 
there is a small contribution to conductivity of \tif~in deeper lying layers 
in the \sto-host whose $d$-bands now slightly overlap the Fermi level. 
Hence it is the lattice relaxations that result in a metallic heterostructure 
and a longer 
healing length towards bulk behavior, in agreement with the experimental 
observations~\cite{ohtomo2002,hwang2006} in spite of a majority of the charge
being tied up at the IF. 
Still, the CO/OO arrangement remains; 
it is robust with respect to relaxation and tetragonal distortion.

Now we summarize.
While the behavior of the many superlattices that we have studied 
produce robust results for the IFs which is easily understood, 
they have a charge- and orbital-ordered character that was
unanticipated from the original reports on these heterostructures.  
If the interaction strength $U$ within the Ti $3d$ states is large enough to
reproduce the AFM insulating state in cubic LTO, then it is more
advantageous for the local charge
imbalance to be accommodated within the IF layer itself, which can be
accomplished by disproportionation, followed by charge order with
\tith~and \tif~distributed in a checkerboard manner.  The interface 
layer is orbitally-ordered, with an occupied  $d_{xy}$-orbital; this
symmetry breaking is due to the intrinsic IF symmetry and is 
unaffected by atomic relaxation.  Indeed both disproportionation and
orbital ordering are insensitive to relaxation.
For the ideal structure, the CO/OO state is a very narrow gap insulator. 
In agreement with previous studies,~\cite{hamann,spaldin06} coupling to the 
lattice is however found to be important in some respects.  
Most notably,
atomic relaxation at the IF shifts the \tith~lower Hubbard band
upward just enough to lead to conducting behavior, which also implies 
a longer healing length towards bulk behavior, consistent with the 
experimental indications.

We acknowledge discussions with 
J. Rustad and J. Kune\v{s}, and communication with A. J. Millis and N. A.
Spaldin.  R.P. was supported through DOE grant 
DE-FG02-04ER15498. W.E.P. was supported by DOE grant DE-FG03-01ER45876 and 
the DOE Computational Materials Science Network.  W.E.P also acknowledges 
support from the Alexander von Humboldt Foundation, and hospitality of
the Max Planck Institute Stuttgart and IFW Dresden, during the latter stages
of this work.


\begin{thebibliography}{10}
\bibitem{details}
For the former cell 18 $k$-points in
 the irreducible part of the Brillouin zone were used.  The
 sphere radii were, in bohr, 2.50 (La,Sr), 1.90 (Ti), 1.60 (O).
Unoccupied La $4f$ states may lie in the same energy
range as the Ti $3d$ states. Although both orbitals are
quite localized and there is no hybridization between
them, we have used $U_{\rm La}=7.5$~eV
to shift the empty $f$-states to higher energies.
\end{thebibliography}
\end{document}